\documentclass[fleqn,usenatbib]{mnras}

\usepackage{newtxtext,newtxmath}

\usepackage[T1]{fontenc}

\DeclareRobustCommand{\VAN}[3]{#2}
\let\VANthebibliography\thebibliography
\def\thebibliography{\DeclareRobustCommand{\VAN}[3]{##3}\VANthebibliography}

\newcommand{\gaia}{\textit{Gaia}}

\usepackage{graphicx}	
\usepackage{amsmath}	

\title[Halo Planets Search]{Exoplanets in ancient stellar populations: occurrence constraints and hot-Jupiter candidates in the Galactic halo}

\author[Bashi, Kunimoto et al.]{
Dolev Bashi$^{1}$\thanks{E-mail: db975@cam.ac.uk}\thanks{These authors contributed equally to this work.},\,
Michelle Kunimoto$^{2}$\footnotemark[2]\thanks{E-mail: mkuni@phas.ubc.ca},\,
Kevin K. Hardegree-Ullman$^{3}$,\,
Tianjun Gan$^{4,5}$,\,
\newauthor
Sharon X. Wang$^{6}$,\,
Zhen Yuan$^{7,8}$\\
$^{1}$Astrophysics Group, Cavendish Laboratory, University of Cambridge, JJ Thomson Avenue, Cambridge CB3 0HE, UK\\
$^{2}$Department of Physics and Astronomy, University of British Columbia, 6224 Agricultural Road, Vancouver, BC V6T 1Z1, Canada\\
$^{3}$Caltech/IPAC-NASA Exoplanet Science Institute, 1200 E. California Boulevard, MC 100-22, Pasadena, CA 91125, USA\\
$^{4}$ Instituto de Astrof\'isica de Canarias (IAC), V\'ia L\'actea s/n, E-38205 La Laguna, Tenerife, Spain\\
$^{5}$ Dept. Astrof\'sica, Universidad de La Laguna (ULL), E-38206 La Laguna, Tenerife, Spain\\
$^{6}$Department of Astronomy, Tsinghua University, Beijing 100084, People’s Republic of China\\
$^{7}$School of Astronomy and Space Science, Nanjing University, Nanjing 210093, People’s Republic of China\\
$^{8}$Key Laboratory of Modern Astronomy and Astrophysics (Nanjing University), Ministry of Education, Nanjing 210093, People’s Republic of China
}

\date{Accepted XXX. Received YYY; in original form ZZZ}

\pubyear{\the\year{}}

\begin{document}
\label{firstpage}
\pagerange{\pageref{firstpage}--\pageref{lastpage}}
\maketitle

\begin{abstract}
The Galactic halo preserves a record of the Milky Way's earliest assembly and contains both in-situ stars and stars accreted from dwarf galaxies. Possible planets around these stars, therefore, probe formation in ancient, metal-poor environments, including systems of extragalactic origin. We present a search for short-period transiting planets around kinematically selected halo dwarfs using \textit{Gaia} DR3 and \textit{TESS}, focusing on planets with periods of $1 < P < 10$ days. We identify two hot-Jupiter (HJ) candidates, one in the in-situ and one in the accreted halo, although the latter is highly grazing and excluded from the occurrence analysis. The accreted candidate, if confirmed, would orbit the most metal-poor HJ host known ([Fe/H] $\approx -1$). Using injection--recovery tests and automated vetting, we constrain occurrence in the full halo, in-situ, and accreted samples. In the HJ regime ($8\,R_\oplus < R_{\rm p} < 22\,R_\oplus$, $1\,\text{day}\ < P < 10$ days), the non-grazing candidate implies an overall halo occurrence rate of  $0.13^{+0.12}_{-0.07}\%$ if planetary, while the absence of confirmed detections gives a corresponding $1\sigma$ upper limit of $<0.14\%$. For the in-situ halo, we infer $0.17^{+0.17}_{-0.10}\%$ (or $<0.19\%$ assuming no detections), while for the accreted halo we derive an upper limit of $<0.56\%$. These rates lie well below the corresponding short-period giant-planet occurrence measured in the Galactic disc. A forward model assuming \textit{Kepler}-like occurrence also predicts $10 \pm 3$ detections compared with at most one observed. We find no significant occurrence difference between the in-situ and accreted halo populations, strengthening the evidence that close-in giant planets are rare across the old, metal-poor halo.
\end{abstract}




\section{Introduction}
\label{sec:intro}

The Galactic halo is composed mainly of metal-poor stars and serves as a key record of our Galaxy’s assembly \citep{Belokurov18, Helmi18, Yuan20, BelokurovKravtsov23}. Its stellar populations include both stars formed \emph{in-situ} during the early stages of the Milky Way and stars later \emph{accreted} from dwarf galaxies. Searching for exoplanets around halo stars, especially those of extragalactic origin, provides a way to test planet formation in low-metallicity environments where it may be suppressed \citep{Boley24}.

In contrast, the Galactic disc, particularly the thin disc, which is characterised by relatively young ages and near-solar metallicities \citep{Bensby03, Adibekyan12}, is planet-rich, with thousands of confirmed planets orbiting disc stars. Large \textit{Kepler} and \textit{TESS} \citep{Ricker15} occurrence rate studies \citep[e.g.,][]{Petigura18, Hsu19, BashiZucker22, KunimotoMatthews20, Kunimoto22, Zink23, Gan23, Bryant23} show that almost all of these detections involve FGKM dwarfs with solar-like metallicities, underscoring how little is currently known about planet formation in the metal-poor halo, where no confirmed exoplanets have been detected in the halo to date.
\citet{Boley21} put the first tight constraint on halo hot-Jupiters (HJs) by searching $\sim10^{4}$ \textit{TESS} halo dwarfs and finding none, yielding a $1\sigma$ upper limit of $0.18\%$ for $P<10$\,d gas giants, an order of magnitude below the $\sim1\%$ rate in the thin disc. A complementary search of $1{,}080$ metal-poor red giants by \citet{Yoshida22} likewise reported zero planets and a $<0.52\%$ (95 \%) limit on similar HJs, while \citet{BashiZucker22} could classify only eleven bona-fide halo targets in their \textit{Kepler} subset, leaving small-planet occurrence essentially unconstrained. Collectively, these non-detections indicate that short-period planets are substantially rarer in the metal-poor halo than in the solar-metallicity disc, although the conclusions are still driven by small-number statistics rather than a genuine census. Establishing robust upper limits or securing the first detections would clarify whether metallicity, formation environment, or the halo’s high binary fraction \citep{Bashi24} dominates planetary outcomes. Moreover, splitting the halo into in-situ and accreted subpopulations allows us to test whether planet formation proceeds differently in dwarf galaxy progenitors versus the early Milky Way.

In this work, we exploit \textit{Gaia} DR3 \citep{Vallenari23, Katz23} kinematics to select nearby halo stars and use \textit{TESS} light curves to search for short-period transiting planets ($P \leq 10 $ days). Section~\ref{sec:sample} outlines the sample selection; Section~\ref{sec:methods} describes the transit search and injection–recovery completeness analysis; Section~\ref{sec:results} presents the resulting occurrence-rate limits, both for the full halo sample and for the in-situ versus accreted subgroups. Section~\ref{sec:discussion} discusses the implications of our results.

\section{Halo Star Sample Selection}
\label{sec:sample}

The halo sample was constructed from \gaia~DR3 \citep{Vallenari23} astrometry and radial velocities \citep{Katz23}, targeting stars with halo-like kinematics in the solar neighbourhood. We applied standard quality cuts to ensure reliable 6D phase-space information, requiring positive parallaxes with (\texttt{parallax\_over\_error > 5}) and well-measured radial velocities (\texttt{rv\_visibility\_periods\_used > 5}, \texttt{rv\_expected\_sig\_to\_noise > 5}). To limit contamination from unresolved binaries, we excluded sources with large radial-velocity uncertainties (\texttt{radial\_velocity\_error > 10}~km~s$^{-1}$, \citealt{Bashi24}) and poor astrometric fits (\texttt{ruwe > 1.4}; \citealt{Lindegren18}).

\begin{figure}
	\includegraphics[width=\columnwidth]{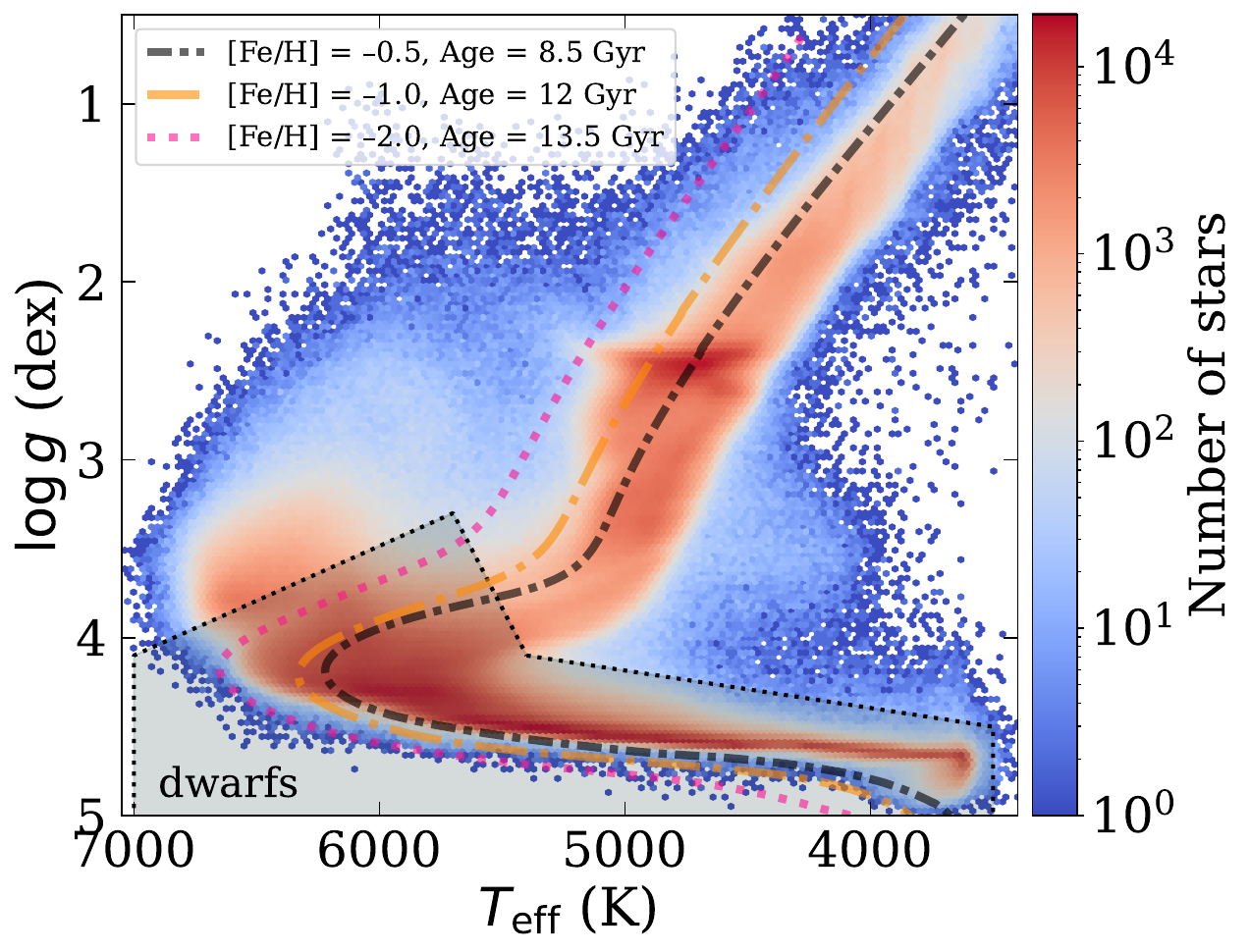}
    \caption{Density plot of effective temperature $T_{\mathrm{eff}}$ versus surface gravity $\log g$ for the full \gaia~DR3 sample considered in this work. The black dotted line and shaded polygon region delineate the dwarf-star selection. Thick dashed and dotted curves show representative metal-poor, old MIST evolutionary tracks typical of halo populations.}
    \label{fig:kiel}
\end{figure}

We restricted the sample to dwarf stars using a Kiel diagram in $(T_{\mathrm{eff}}, \log g)$, based on stellar parameters derived from the \gaia~low-resolution XP spectra of \citet{Andrae23}. Figure~\ref{fig:kiel} shows the density distribution of the full sample ($14{,}823{,}507$ sources), with the adopted polygon selection\footnote{Defined by the vertices ($T_{\mathrm{eff}}$, $\log g$): [(5,400, 4.1), (3,500, 4.5), (3,500, 5.4), (7,000, 5.35), (7,000, 4.1), (5,700, 3.3), (5,400, 4.1)] in units of K and dex, respectively.} isolating dwarf stars ($7{,}748{,}323$ sources). The selection is validated by comparison with metal-poor, old MIST isochrones \citep{MESA,MIST0,MIST1}, appropriate for typical halo populations.

Galactic space velocities $(U,V,W)$ were computed using \texttt{galpy} \citep{Bovy15} and corrected to the Local Standard of Rest assuming $(U,V,W)_\odot=(11.1,12.24,7.25)$~km~s$^{-1}$ \citep{Schonrich10}. Orbital energies $E$ and angular momenta $L_z$ were calculated in the \citet{McMillan17} axisymmetric Galactic potential. 

Halo candidates were selected by requiring a total Galactocentric velocity $V_{\mathrm{tot}}=\sqrt{U^2+V^2+W^2} > 220$~km~s$^{-1}$, a commonly adopted threshold that separates the non-rotating halo from the disc population in the solar neighbourhood \citep[e.g.,][]{NissenSchuster10,Helmi18,Belokurov18,BashiZucker22}.

We cross-matched the halo sample with the \textit{TESS} Input Catalogue (TIC). Source positions were propagated from the \gaia~DR3 reference epoch (J2016) to J2000, and a $1''$ cone search was applied, yielding $20\,836$ matched sources.

To distinguish between accreted and in-situ halo populations, we adopted the kinematic classification of \citet{BelokurovKravtsov23} in the $E$–$L_z$ plane. Because orbital energy depends on the adopted Galactic potential, we recalibrated their separation curve to the \citet{McMillan17} potential by aligning the solar orbital energy, applying a uniform shift of $\Delta E \simeq 0.242\times10^5$~km$^2$~s$^{-2}$. Stars above the adjusted boundary with $L_z < 0.58\times10^3$~kpc~km~s$^{-1}$ were classified as accreted, while the remainder were labelled in-situ. 

Here, `accreted' should be understood as an operational label for sources selected by the accreted region of the \citet{BelokurovKravtsov23}
classification, which was shown to separate the chemically calibrated in-situ and accreted populations with high accuracy, $\gtrsim 95$ per cent \citep{BelokurovKravtsov24}. At the same time, this assignment is necessarily statistical: in-situ or heated-disc stars can overlap with accreted debris in dynamical space, and simulations suggest that the contamination of specific halo substructure selections can in some cases reach tens of per cent \citep[e.g.][]{Thomas25}. We therefore interpret the accreted/in-situ division at the population level, rather than as a definitive birth-origin assignment for every individual source. Our conservative
selection preserves the qualitative shape of the \citet{BelokurovKravtsov23} boundary while avoiding the region most strongly populated by the high-$\alpha$ disc.


\begin{figure*}
	\includegraphics[width=14cm]{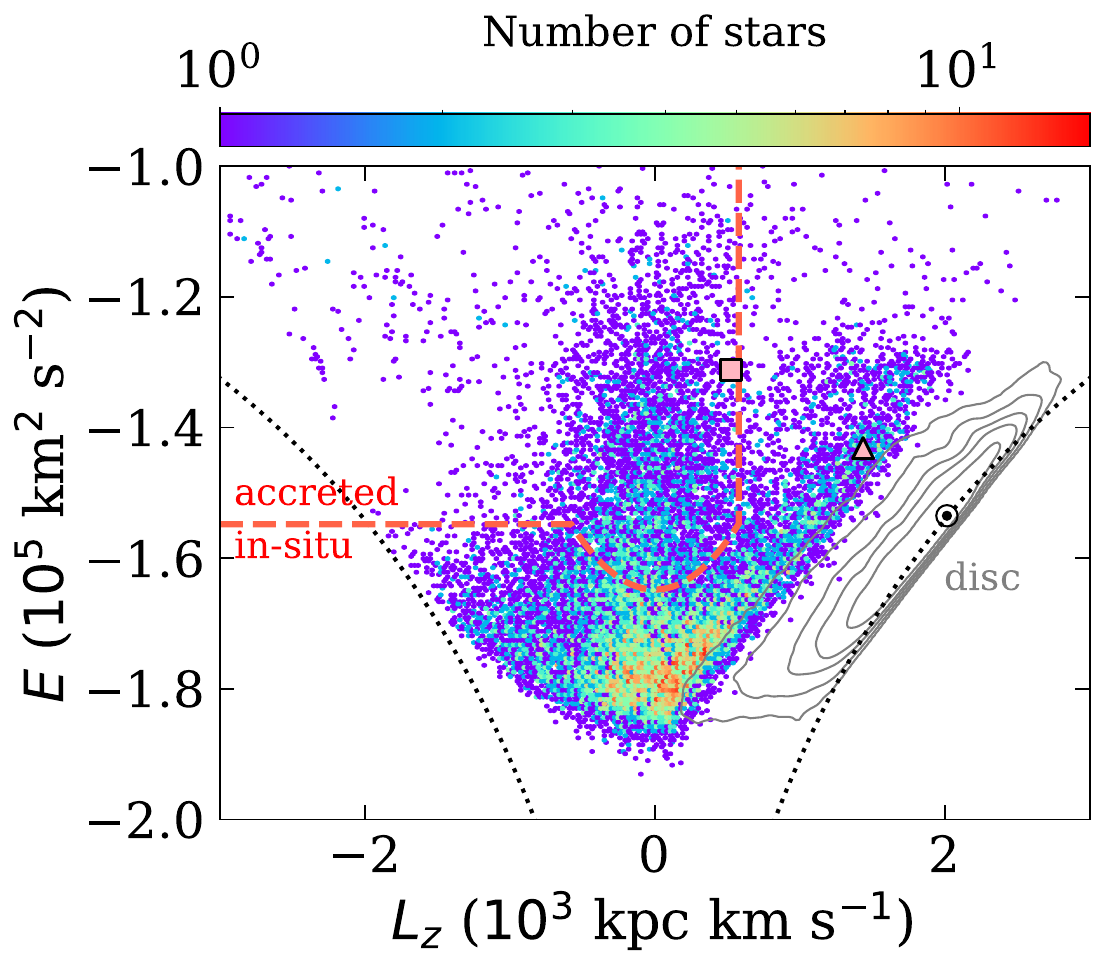}
\caption{
Energy–angular momentum ($E$–$L_z$) distribution for the kinematically selected halo sample. The separation curve between accreted and in-situ populations is marked by a red dashed curve (see text for further details). Stars above the curve with $L_z < 0.58\times10^3$\,kpc\,km\,s$^{-1}$ are classified as accreted, while those below or with higher $L_z$ are in-situ. Grey contours indicate disc stars excluded by the halo kinematic selection. This division is used throughout to compare planet occurrence in accreted and in-situ halo populations. The black fine-dashed curve corresponds to the maximal angular momentum at fixed energy, i.e. orbits with circular velocity. $\odot$ marks the location of the Sun in the chosen potential. Pink markers show our two substellar host candidates (see Table~\ref{tab:PCs} for more information).
}
    \label{fig:E_Lz}
\end{figure*}

Figure~\ref{fig:E_Lz} shows on the $E-L_z$ plane a density plot of the halo dwarf sample. Red dashed line delineates our selection cut of in-situ and accreted stars. To guide the eye, we also added grey contours of the disc sample of stars. The samples are confined by a dotted black curve, which is the bounding parabola defined by the escape velocity as a function of the change in energy at a fixed Galactocentric radius. The Sun's location is marked in black at $E_{\odot}=-1.53 \times 10^5~\mathrm{km^2~s^{-2}}$. As can be clearly seen, the solar neighbourhood is mainly populated by disc stars. 

\begin{figure}
	\includegraphics[width=\columnwidth]{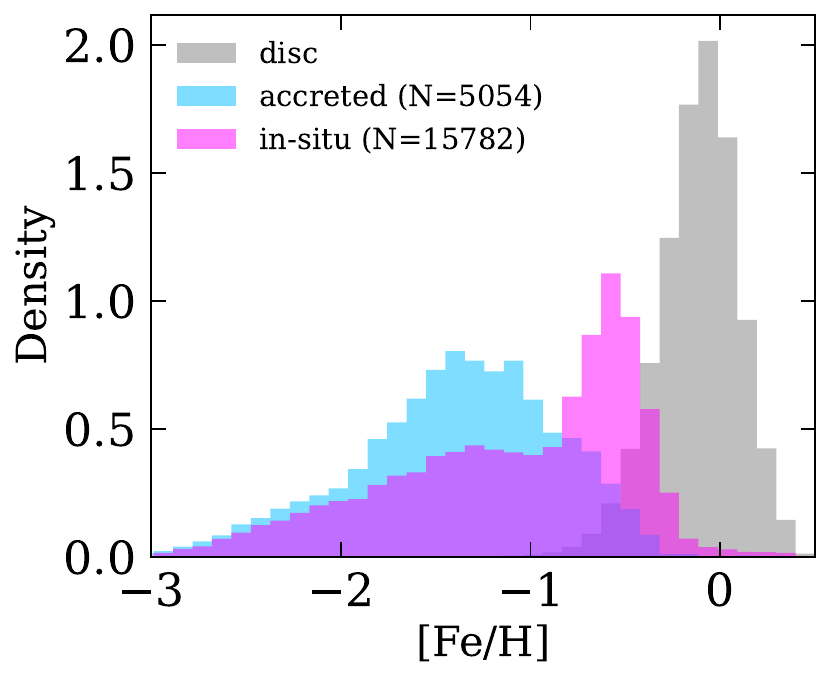}
    \caption{A normalised histogram of metallicity [Fe/H] using metallicities derived from the \gaia\ XP spectra via the XGBoost model of \citet{Andrae23}. The affiliation of halo stars to accreted (blue) and in-situ (magenta) is based on the separation curve shown in Figure~\ref{fig:E_Lz}. For reference, we have added the distribution of disc stars in grey.}
    \label{fig:feh_hist}
\end{figure}

After applying all selection criteria, the final halo sample consists of $5{,}054$ accreted and $15{,}782$ in-situ stars. Figure~\ref{fig:feh_hist} shows the corresponding [Fe/H] distributions, which are consistent with expectations: the accreted population (blue histogram) is predominantly metal-poor, while the in-situ sample (magenta histogram) spans both the high-$\alpha$ disc component (right peak in magenta histogram) and a metal-poor tail, possibly associated with the \emph{Aurora} population \citep{BelokurovKravtsov23}.

For our analysis, we adopted homogeneous stellar parameters derived using the photometric classification pipeline of \citet{HardegreeUllman20,HardegreeUllman25}. This approach combines \gaia\ and 2MASS photometry with machine-learning–based metallicity estimates and MIST stellar models \citep{MIST0,MIST1} to infer $T_{\mathrm{eff}}$, $\log g$, [Fe/H], stellar radius, and mass. In particular, we used the XGBoost-derived atmospheric parameters from \citet{Andrae23}, whose metallicities are trained on an APOGEE-based spectroscopic sample supplemented with metal-poor stars. These parameters are internally consistent and well-suited for metal-poor populations, providing a homogeneous set of stellar properties for our halo sample. Full methodological details of the parameter inference framework are given in \citet{HardegreeUllman25}. 

\section{Planet search process}
\label{sec:methods}

\subsection{Light Curves}

Of the $20{,}836$ halo stars, $11{,}190$ stars ($8{,}389$ in-situ and $2{,}801$ accreted) have \textit{TESS} Full-Frame Image (FFI) light curves produced by the Quick-Look Pipeline \cite[QLP;][]{QLPPM,QLPEM1p1,QLPEM1p2} from Sectors 1 -- 93. We chose the QLP dataset as it provides the longest coverage across multiple \textit{TESS} sectors compared to other publicly available datasets. For each light curve, we removed cadences with non-zero quality flags (indicating poor quality), and normalised the fluxes to 1 before combining individual sectors into a multi-sector light curve. We then detrended each light curve using the biweight time-windowed slider implemented in the \texttt{wotan} package \citep{Hippke2019} with a window of 0.5 days to remove astrophysical variability. Finally, we $5\sigma$-clipped outliers in the data. Only outliers in the positive direction were removed so as to leave any deep transits untouched.

\subsection{Transit Candidate Identification}

We searched the light curves for signals with periods $1 < P < 10$ days using a box-fitting least squares (BLS) routine \citep{Kovacs2002} implemented in the \texttt{cuvarbase} time series analysis package \citep{Hoffman2022}. We identify an initial detection based on the period with the strongest power in the BLS periodogram. However, because BLS occasionally recovers an orbital period that is an integer factor of the correct period, we then performed a series of trapezoid model fits with periods initialised to 1/3, 1/2, 1, 2, and 3 times the BLS-detected period, and determined the properties of the final potential transit detection based on the fit resulting in the largest signal-to-noise ratio (SNR). 

The trapezoid model fits are parameterised by the signal's orbital period, transit epoch ($T_{0}$), transit depth ($\delta$), ratio between full transit duration and orbital period ($q$), ratio between ingress duration and full transit duration ($q_{\mathrm{in}}$), and out-of-transit flux level ($f_{0}$). Following \citet{Kipping2023}, the signal-to-noise ratio (SNR) of a trapezoidal transit is
\begin{equation}\label{eqn:SNR}
    \mathrm{SNR} = \sqrt{\frac{1 + 2 T_{23}/T_{14}}{3}} \bigg(\frac{\delta}{\sigma_{tr}}\bigg)
\end{equation}
where $T_{14} \equiv qP $ is the transit duration, $T_{23} \equiv qP(1 - 2q_{\mathrm{in}})$ is the totality duration, and $\sigma_{tr}$ is the noise over the transit duration. The term in parentheses is the most commonly quoted definition of transit SNR, which assumes a box shape, but this does not capture reductions to signal-to-noise ratio due to grazing transits as accurately as the trapezoid model.

In order to account for correlated noise in the \textit{TESS} light curves, we adopt the pink noise definition from \citet{Pont2006} to estimate $\sigma_{tr}$ as
\begin{equation}\label{eqn:noise}
    \sigma_{tr} = \sqrt{\frac{\sigma_{w}^{2}}{n} + \frac{\sigma_{r}^{2}}{N_{tr}}},
\end{equation}
where $\sigma_{w}$ and $\sigma_{r}$ represent the white and red noise components, respectively, $n$ is the number of in-transit data points, and $N_{tr}$ is the number of observed transits. The white- and red-noise terms are computed following the procedure described by \citet{HartmanBakos2016} and \citet{LEOVetter} for estimating pink noise in time-series photometry.

We define Threshold Crossing Events (TCEs) as signals with SNR $> 9$ and $N_{tr} \geq 3$. A total of $2{,}025$ signals met this detection criterion. Among these TCEs, we recovered all four known \textit{TESS} Objects of Interest with periods $1 < P < 10$ days (TOI-152.01, TOI-324.01, TOI-5962.01, and TOI-6235.01).

Because the original detrending process can be destructive to the shape of a planetary transit and often results in a loss in SNR, we re-detrended each light curve with the TCEs masked out before proceeding with further candidate vetting. Unless otherwise specified, these re-detrended light curves are used for all later analyses in this paper.

\subsection{Automated Candidate Vetting}

Vetting of TCEs is necessary to distinguish planets from the enormous number of false alarms (FAs; e.g., instrumental systematics, intrinsic stellar variability) and astrophysical false positives (FPs; e.g., bound eclipsing binaries, nearby eclipsing systems). Doing so in a fully automated, reproducible way is also needed to produce planet catalogues suitable for statistically robust occurrence rate studies; any introduction of manual inspection will introduce uncharacterisable biases into the planet detection process.

We used \texttt{LEO-Vetter} \citep{LEOVetter} to identify the most promising planet candidates (PCs) from our sample of 2,025 TCEs. Inspired by the \textit{Kepler} Robovetter \citep{Coughlin2016,Thompson2018}, \texttt{LEO-Vetter} computes a set of metrics at both the flux- and pixel-level, which are then compared to pass-fail thresholds to determine which signals are PCs (TCEs that pass all tests) and which are FAs or FPs (TCEs that fail at least one test). Flux-level tests include those that look for transit asymmetries, non-uniqueness, odd-even transit depth differences, and significant secondary eclipses, while pixel-level vetting involves producing difference images to look for centroid offsets. A total of five PCs passed both flux- and pixel-level vetting, including two TOIs (TOI-5962.01, and TOI-6235.01) and three signals which are not yet known TOIs (TIC-7262939.01, TIC-263176654.01, and TIC-421991589.01). 

The two TOIs that did not pass \texttt{LEO-Vetter} were TOI-152.01 and TOI-324.01. TOI-152.01 was flagged as a nearby eclipsing binary (NEB) by pixel-level vetting, and is also considered an NEB by the \textit{TESS} Follow-Up Observing Program Working Group (TFOPWG), as noted on ExoFOP \citep{ExoFOP}. TOI-324.01 was flagged as being too large ($R_{p} > 22~R_{\oplus}$) by flux-level vetting. This TOI is still an active planet candidate, though a comment in the TOI catalogue suggests that it has a $\sim2300$ ppm secondary visible in the light curve from NASA's Science Processing Operations Centre (SPOC), which is consistent with an eclipsing binary.

Of the candidates that passed \texttt{LEO-Vetter}, TOI-5962.01 has already been shown to be a false positive based on TFOPWG ground-based photometric follow-up; specifically, it is a nearby planet candidate (NPC). According to \citet{Zhang2026}, the correct host star is a fainter binary star companion only $1.7^{\prime\prime}$ away, which was too closely separated to be distinguished by pixel-level vetting. The \textit{Gaia} astrometry of the two sources is consistent with a physically associated common-proper-motion binary system, with a projected separation of roughly $\sim 200$ AU. 
\citet{Zhang2026} estimated a dilution-corrected planet radius of $\sim8.3~R_{\oplus}$ for the eclipsing object around the companion, making it either a genuine giant planet, a brown dwarf, or a third star in the system. While it may still be planetary, we removed this object from our planet candidate catalogue as it does not orbit one of the stars in our sample.

To both further remove false positives and compute final planet properties for the four surviving PCs, we used \texttt{pymc} to fit transit models implemented in the \texttt{exoplanet} package \citep{Foreman-Mackey21}. \texttt{LEO-Vetter} performs a simple least-squares transit model fit to estimate planet radius from the planet-to-star radius ratio ($R_{p}/R_{s}$) and stellar radius, among other metrics; however, it is recommended that users perform more comprehensive transit model fits to derive publishable planet parameters, especially because grazing transits tend to have their planet radii underestimated by this fitting routine due to capping impact parameter ($b$) at 1 \citep{LEOVetter}. 

We sampled four chains for 2000 tuning steps and 2000 posterior draws per chain, and confirmed that the chains converged according to the Gelman-Rubin convergence statistic for each parameter satisfying $\hat{r} < 1.01$ \citep{Gelman1992}. We let $P$, $T_{0}$, $R_{p}/R_{s}$, and $b$ vary as fit parameters. $M_{s}$, and $R_{s}$ were fixed to their stellar catalogue values \citep{HardegreeUllman20, HardegreeUllman25}. We adopted a quadratic limb-darkening law parameterised by $q_{1}, q_{2}$ from \citet{exoplanet:kipping13}, and fit for flux offsets and jitter terms added in quadrature. We propagated the uncertainty from $R_{s}$ and $R_{p}/R_{s}$ when computing the uncertainty on $R_{p}$. 

As a result of this additional fitting step, TIC-7262939.01 ($R_{p} = 24.20_{-10.69}^{+37.08}~R_{\oplus}$) and TOI-6235.01 ($R_{p} = 26.91_{-8.41}^{+14.39}~R_{\oplus}$) became too large to be planetary in addition to having high impact parameters ($b > 1$), and thus we disposition them as FPs. TOI-6235.01 is currently considered an Ambiguous Planet Candidate (APC) by the TFOPWG due to a $\sim25~{\mathrm{kms^{-1}}}$ radial velocity semi-amplitude, which is consistent with the interpretation that this is indeed an eclipsing binary star. 

The fitted transit model parameters are shown in Table \ref{tab:fits}, along with comments summarising our conclusions about each PC.

\begingroup
\setlength{\tabcolsep}{6pt}     
\renewcommand{\arraystretch}{1.5} 

\begin{table*}
    \centering
    \begin{tabular}{ccc|cccc|c}
    \hline\hline
        Sample & TIC ID & Comments & $R_{p}/R_{s}$ & $P$ (days) & $T_{0}$ (BJD - 2457000) & $b$ & $R_{p}$ ($R_{\oplus}$)\\        
    \hline
        in-situ & 150902766 & too large; TOI-6235.01; & $0.303_{-0.095}^{+0.161}$ & $7.82061_{-0.00032}^{+0.00030}$ & $2856.7270_{-0.0016}^{+0.0016}$ & $1.16_{-0.11}^{+0.18}$ & $26.91_{-8.41}^{+14.39}$ \\
        & & spectroscopic binary & & & & & \\
        in-situ & 421991589 & planet candidate & $0.1521_{-0.0046}^{+0.0054}$ & $5.001197_{-0.0000082}^{+0.0000081}$ & $2452.4097_{-0.0014}^{+0.0015}$ & $0.776_{-0.045}^{+0.033}$ & $13.29_{-0.53}^{+0.59}$\\
        accreted & 7262939 & too large & $0.195_{-0.086}^{+0.301}$ & $2.306403_{-0.0000045}^{+0.0000047}$ & $1385.2153_{-0.0020}^{+0.0021}$ & $1.06_{-0.11}^{+0.11}$ & $24.20_{-10.69}^{+37.08}$ \\
        accreted & 263176654 & planet candidate & $0.212_{-0.107}^{+0.265}$ & $5.104095_{-0.0000058}^{+0.0000053}$ & $1330.0693_{-0.0017}^{+0.0019}$ & $1.14_{-0.12}^{+0.28}$ & $14.82_{-7.51}^{+18.39}$ \\
    \end{tabular}
    \caption{Transit model parameters fitted using \texttt{exoplanet} for the four surviving PCs passing \texttt{LEO-Vetter}. Planet radius is derived from the fitted $R_{p}/R_{s}$ and the stellar radius. Only TIC-421991589.01 and TIC-263176654.01 comprise our final planet catalogue (Table \ref{tab:PCs}), as TIC-150902766.01 and TIC-7262939.01 are too large to be planetary.}
    \label{tab:fits}
\end{table*}
\endgroup

\subsection{Final Planet Catalogue}\label{sec:final}

\begin{figure*}
\centering
\includegraphics[width=0.44\linewidth]{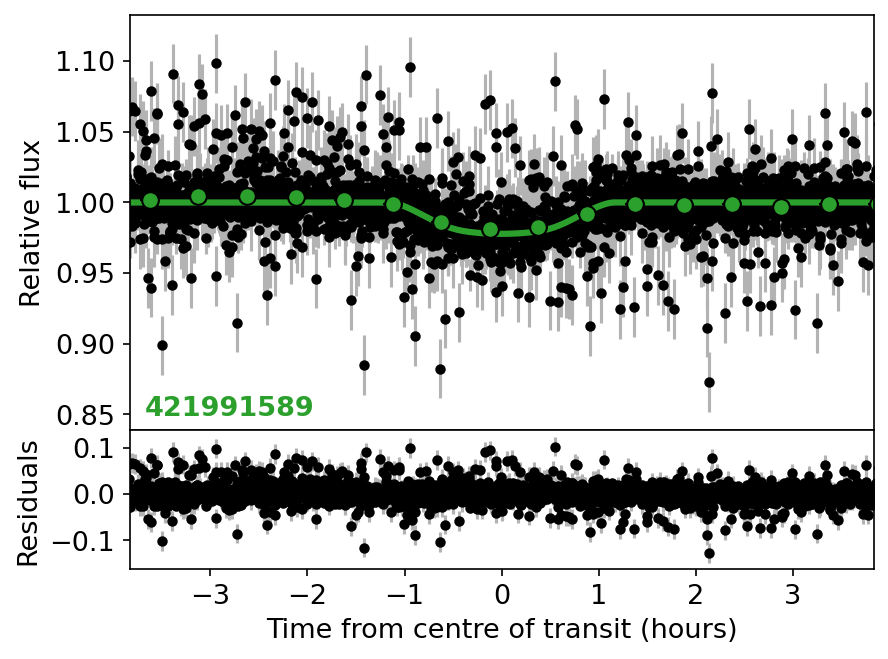}
\includegraphics[width=0.45\linewidth]{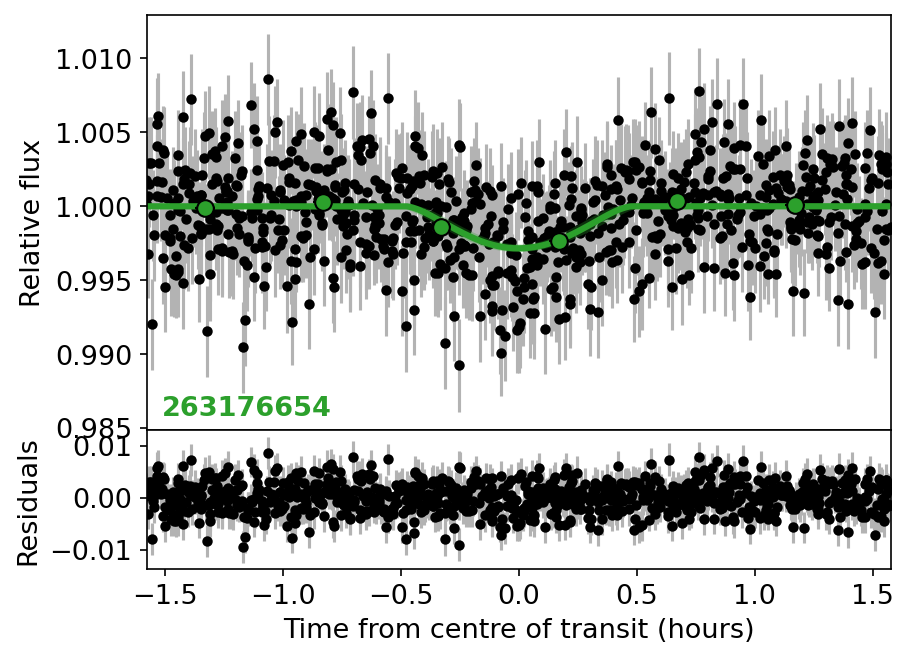}
\caption{\textit{TESS} phase-folded light curves for the two planet candidates recovered in this work. Of these two candidates, only TIC-421991589.01 has a non-grazing orbit ($b < 0.9$), and therefore is the only candidate included in our subsequent occurrence rate calculations (\S\ref{sec:occrates}). The grey points reflect the photometric observations while green coloured circles are data averaged in 30 min bins. The solid lines show the median transit model.}
\label{fig:PCs}
\end{figure*}

Our final planet catalogue contains two PCs, both HJ ($R_{p} > 8~R_{\oplus}$, $P < 10$ days), shown in Table \ref{tab:PCs} and Figure \ref{fig:PCs}: one in the in-situ sample (TIC-421991589.01) and one in the accreted sample (TIC-263176654.01). TIC-421991589.01 ($R_{p} = 13.29_{-0.53}^{+0.59}~R_{\oplus}$, $P \sim 5.0$ days) orbits a high-$\alpha$ (\emph{Splash}) star \citep{Belokurov18} with [Fe/H]~$\approx-0.29$, which is notable given its relatively high-metallicity environment for a halo-like host. Meanwhile, if confirmed, TIC-263176654.01 ($R_{p} = 14.82_{-7.51}^{+18.39}~R_{\oplus}$, $P\sim5.1$ days) will have the lowest-metallicity host star ([Fe/H] $\approx$ -0.93) among HJ hosts known so far. However, it also has a highly grazing geometry ($b = 1.14_{-0.12}^{+0.28}$). We caution that this increases the likelihood that this is an eclipsing binary star.

\begingroup
\setlength{\tabcolsep}{6pt}     
\renewcommand{\arraystretch}{1.5} 

\begin{table*}
    \centering
    \begin{tabular}{ccccccc|cc}
    \hline\hline
        Marker & Sample & Gaia DR3 ID & TIC ID & $R_{s}$ ($R_{\odot}$) & $M_{s}$ ($M_{\odot}$) & [Fe/H] & $R_{p}$ ($R_{\oplus}$) & $P$ (days) \\        
    \hline
        $\huge\triangle$ & in-situ & 2653951865196561920 & 421991589 & $0.802\pm0.02$ & $0.772\pm0.061$ & $-0.288$ & $13.29_{-0.53}^{+0.59}$ &  $5.001197_{-0.0000082}^{+0.0000081}$ \\
        \fbox{\rule{0pt}{0.01cm}\rule{0.01cm}{0pt}} & accreted & 4631496645276178048 & 263176654 & $0.640\pm0.015$ & $0.664\pm0.052$ & $-0.934$ & $14.82_{-7.51}^{+18.39}$ & $5.104095_{-0.0000058}^{+0.0000053}$ \\
    \end{tabular}
    \caption{Stellar and planetary parameters for the transiting planet candidates recovered in this work. Only TIC-421991589.01 is included in the final occurrence rate calculations due to its non-grazing transit geometry ($b < 0.9$; Table \ref{tab:fits}). Marker symbols correspond to those used in Figure~\ref{fig:E_Lz}.}
    \label{tab:PCs}
\end{table*}
\endgroup

\section{Occurrence Rate Methodology}\label{sec:occrates}

\subsection{Pipeline Detection Efficiency}

The characterisation of our pipeline's detection efficiency, i.e., the fraction of planets orbiting the stars in our sample that we expect to have been detected by our pipeline, is crucial for de-biasing our planet catalogue for occurrence rate calculations. As in most previous \textit{TESS} occurrence rate studies \cite[e.g.][]{Boley21,Zhou2019,Fernandes2022,BeleznayKunimoto22}, we characterise detection efficiency at the population level by performing a suite of injection/recovery tests and taking an average over all stars.

We injected twenty transits into each pre-detrended light curves (totalling $\sim224{,}000$ injections), with orbital periods log-uniformly drawn over $P \in (1, 10)$ days, planet radii log-uniformly drawn over $R_{p} \in (2, 22)~R_{\oplus}$, impact parameters uniformly drawn over $b \in (0, 0.9)$, and transit epochs uniformly drawn in orbital phase. Each transit was created using a quadratic limb-darkening model \citep{MandelAgol2002} assuming circular orbits, with limb-darkening coefficients interpolated from \citet{Claret2017} limb-darkening tables. Each injected light curve was then detrended and searched using the same process as for the original data. We consider an injected planet as recovered by BLS if the TCE detection criteria were met and the detected ephemerides matched the injected period and transit epoch with significance $\sigma_{P} > 2.5$ and $\sigma_{T} > 2$, as defined by \citet{Coughlin2014}. 

The vetting procedure will also result in a further reduction in detection efficiency, as \texttt{LEO-Vetter} may not necessarily successfully classify an injection as a planet. We ran \texttt{LEO-Vetter}'s flux-level tests on all injected signals that were recovered by BLS, and measured our overall pipeline detection efficiency as the fraction of injections both successfully recovered by BLS and classified as PCs. Because we injected planets at the light curve level rather than the pixel level, we are unable to characterise further reduction in efficiency due to imperfect pixel-level vetting; however, we assume that this is negligible because \texttt{LEO-Vetter} pixel-level vetting was explicitly designed to favour high completeness \citep{LEOVetter}.

\begin{figure*}
\centering
\includegraphics[width=0.48\linewidth]{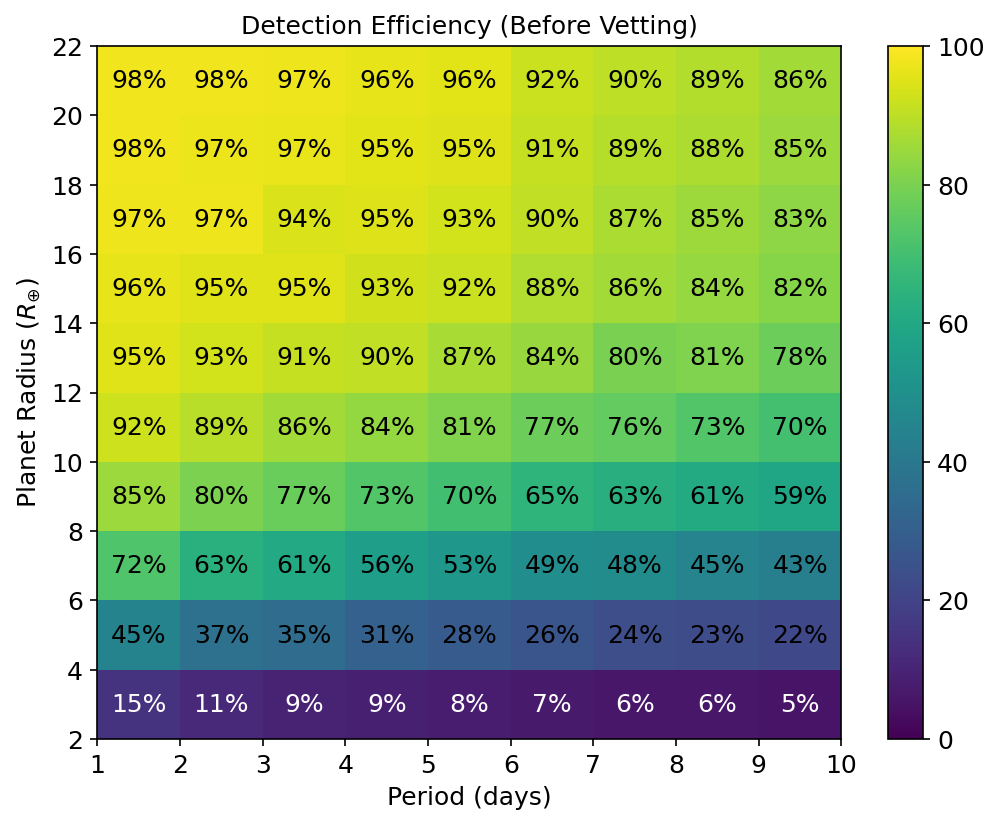}
\includegraphics[width=0.48\linewidth]{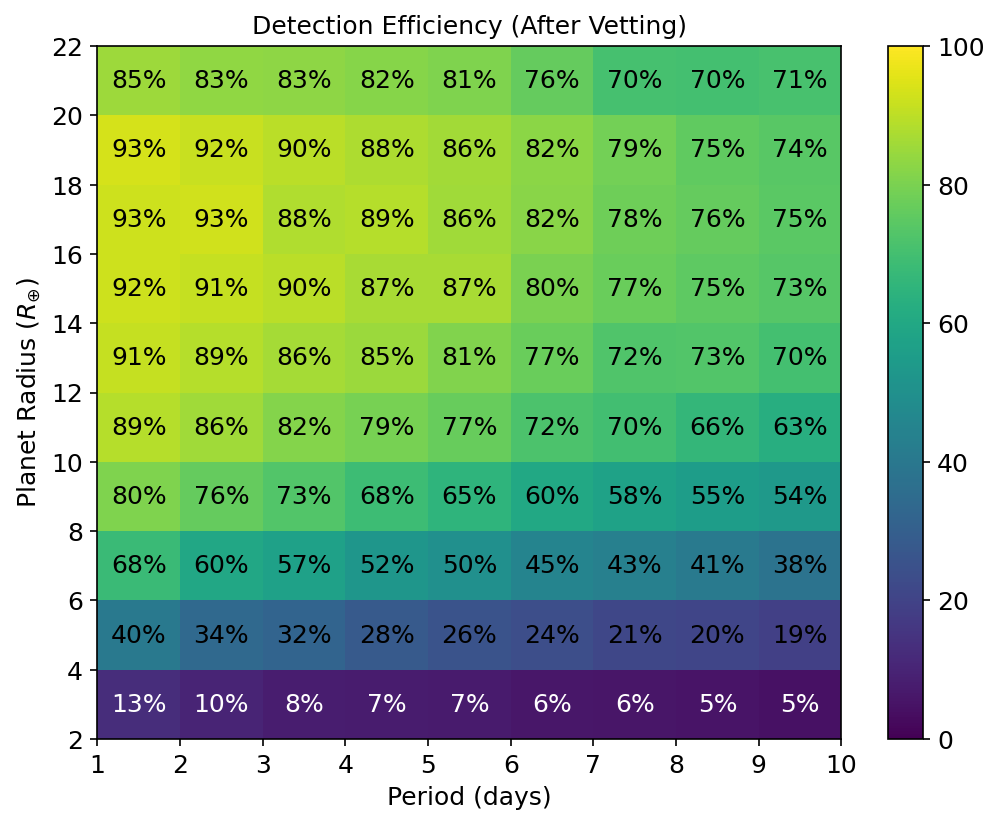}
\caption{The detection efficiency of our pipeline, showing the percentage of transit injections recovered as a function of planet radius and orbital period following only the BLS search process (left) and after the automated vetting process (right), averaged over all $11{,}190$ stars in our sample. While efficiency generally increases at larger radii (and shorter periods), there is a slight reduction for the biggest objects ($R_{p} \gtrsim 20~R_{\oplus})$ due to the vetting process. These planets are very close to \texttt{LEO-Vetter}'s cut on planet size ($R_{p} < 22~R_{\oplus}$) and are harder to distinguish from grazing eclipsing binary stars.}\label{fig:completeness}
\end{figure*}

Our detection efficiency in a coarse $10\times9$ grid in $R_{p}-P$ space is shown in Figure \ref{fig:completeness}, both before and after vetting. We generally have higher completeness at shorter periods and larger planet radii, and recover 87\% of the shortest-period ($P < 3$ days) HJs (92\% recovery if only considering the BLS detection process).

\subsection{Occurrence Rate Calculations}

We compute planet occurrence rates (average number of planets per star) in distinct cells across $\log R_{p}-\log P$ space using a Bayesian methodology originally described by \citet{Hsu2018}. In summary, the planet population is modelled as a Poisson point process in $\log R_{p}-\log P$ space with constant rate within each cell, such that the expected number of planets per star in the cell is $f_{\mathrm{cell}}$. The posterior distribution for $f_{\mathrm{cell}}$ is given by
\begin{equation}
    p(f_{\mathrm{cell}}|n_{\mathrm{pl}}, n_{\mathrm{eff}}) \sim \mathrm{Gamma}(\alpha_{0} + n_{\mathrm{pl}}, \beta_{0} + n_{\mathrm{eff}}),\label{eqn:gamma}
\end{equation}
where $\alpha_{0}$ and $\beta_{0}$ are the shape parameter and rate (inverse scale) parameter that define a gamma distribution prior on $f_{\mathrm{cell}}$, $n_{\mathrm{pl}}$ is the number of planets observed in the cell, and $n_{\mathrm{eff}}$ is the effective number of stars searched. As in \citet{Hsu2018}, we set $\alpha_{0} = \beta_{0} = 1$, which corresponds to a prior on $f_{\mathrm{cell}}$ of $p(f_{\mathrm{cell}}) = \exp(-f_{\mathrm{cell}})$.

If all planets were perfectly detectable, $n_{\mathrm{eff}}$ would be simply equal to the number of surveyed stars ($n_{s}$). In practice, we must account for both non-transiting orbits and the lack of detectability due to insufficient SNR for a given planet size and orbital period. Following \citet{Hsu2018, Bashi20, BeleznayKunimoto22}, we estimate the effective number of stars $n_{\mathrm{eff}}$, for each cell via Monte Carlo. For each star, we draw $N_\mathrm{samp} = 100$ planets (with periods and radii drawn log-uniformly within the cell limits) and estimate the average detection probability, 
\begin{equation}
p_{j} = (1/N_\mathrm{samp})\Sigma_{i=1}^{N_\mathrm{samp}}p_{\mathrm{tr},i,j}p_{\mathrm{det},i,j},
\end{equation}
where $p_{\mathrm{tr},i,j}$ and $p_{\mathrm{det},i,j}$ are the geometric transit probability and pipeline detection and vetting probability for the $i$-th draw of parameters. The final expected value for $n_{\mathrm{eff}}$ is then the sum over all surveyed stars, $n_{\mathrm{eff}} = \Sigma_{j=1}^{n_s}p_{j}$.

For the geometric transit probability, we assume zero eccentricity and use
\begin{equation}
    p_{\mathrm{tr}} = 0.9\frac{(R_{p}+R_{s})}{a},
\end{equation}
where $a$ is the semi-major axis of a given planet's orbit and the factor of 0.9 accounts for the fact that we only consider planets with $b < 0.9$ for our occurrence rates. Planets in more grazing orbits will have unreliable radii and are therefore more likely to be confused for grazing eclipsing binaries.

For pipeline detection probability, we compile our star-averaged detection efficiency in a $30\times30$ grid in $\log R_{p}-\log P$ space and interpolate using scipy's \texttt{RectBivariateSpline} to estimate $p_{\mathrm{det}}$ for any randomly drawn period and radius. Ideally, we could have characterised detection efficiency at the per-target level for this step, but this was infeasible due to the computational expense of our planet search and vetting pipeline. Nevertheless, we believe this choice should have a negligible effect on our occurrence rates because the calculation of $n_{\mathrm{eff}}$ effectively averages over all stars.

\section{Results}
\label{sec:results}

\subsection{Occurrence rates and upper limits in the accreted and in-situ halo samples}

\begin{figure}
    \includegraphics[width=\linewidth]{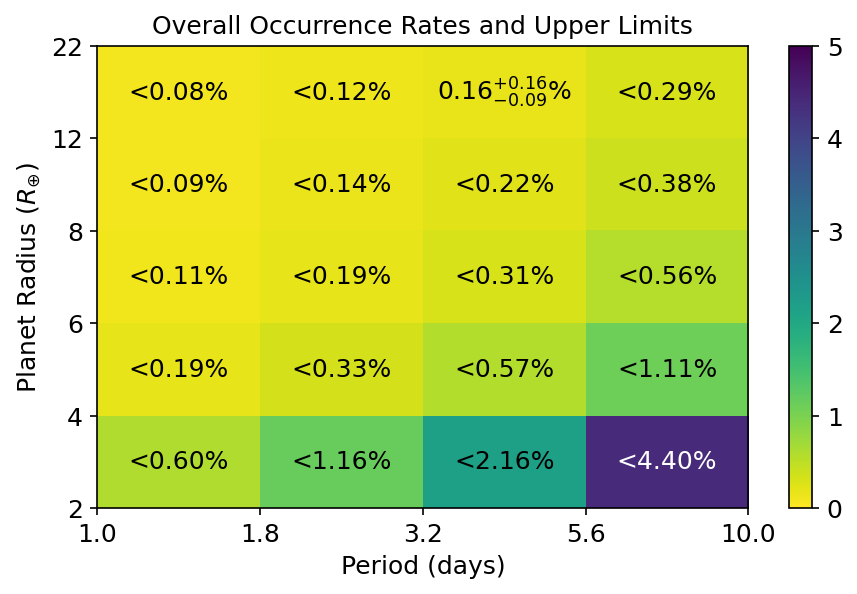}
    \includegraphics[width=\linewidth]{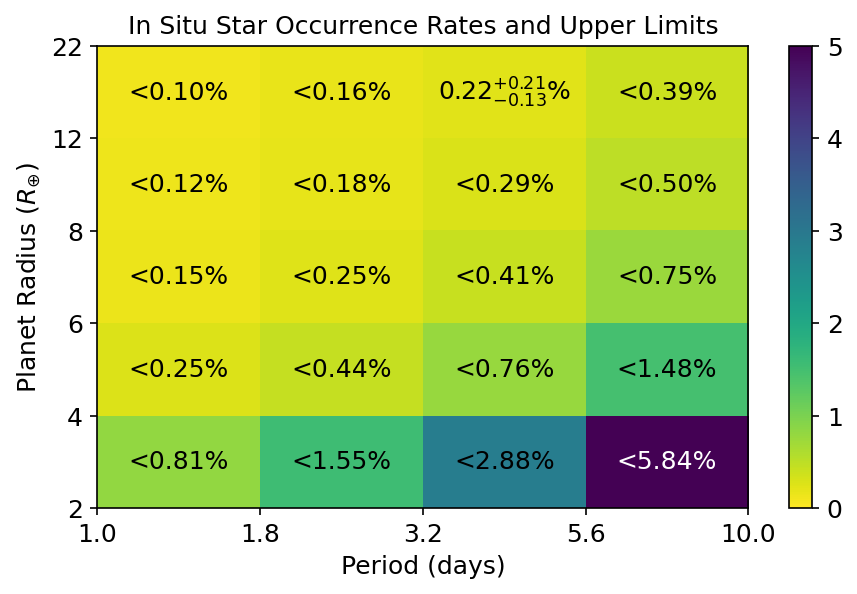}
    \includegraphics[width=\linewidth]{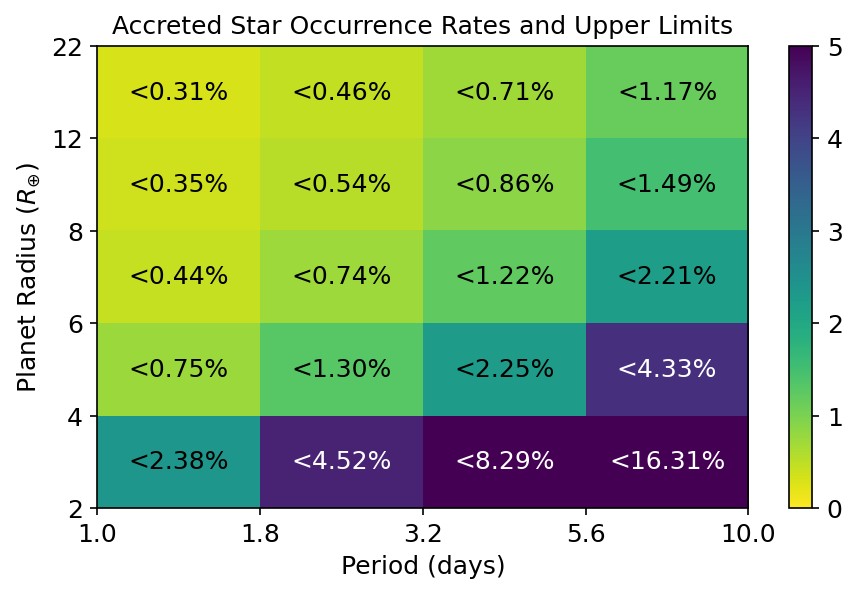}
    \caption{2D occurrence constraints in the period-radius plane for short-period planets ($P<10$\,days) around halo dwarfs, shown for (top) the full halo sample, (middle) the in-situ halo, and (bottom) the accreted halo. Most cells have zero planet detections, for which we report only $1\sigma$ upper limits. These maps highlight the paucity of close-in planets in the halo.}\label{fig:occrates}
\end{figure}

With only one planet candidate with $b < 0.9$ (TIC-421991589.01) in our catalogue, our occurrence rates will largely be upper limits. We report $1\sigma$ upper limits \cite[following the previous \textit{TESS} halo star occurrence rate work;][]{Boley21} based on the 68.3\% credible interval of the Gamma distribution (Eq. \ref{eqn:gamma}) with $n_{\mathrm{pl}} = 0$. For the cell containing TIC-421991589.01, we report the median of the Gamma distribution with lower and upper uncertainties given by the 68.3\% credible interval with $n_{\mathrm{pl}} = 1$. These occurrence rates are shown in Figure \ref{fig:occrates} for the entire 11,190-star QLP sample, and further split into in-situ and accreted subsets. We caution that our candidate is unconfirmed, and that with only zero to one detections we are in a fundamentally low-information regime.

Averaging over the entire HJ parameter space ($8 < R_{p} < 22~R_{\oplus}$, $1 < P < 10$ days), we find an overall occurrence rate of $0.13_{-0.07}^{+0.12}\%$, assuming TIC-421991589.01 is a bona fide planet (alternatively, a $1\sigma$ upper limit of $<0.14\%$ if this object is a false positive). This abundance is significantly lower than the $\sim0.5\%-0.6\%$ occurrence rates derived for Sun-like stars observed by \textit{Kepler} \cite[e.g.][]{Petigura18,KunimotoMatthews20} and \textit{TESS} \citep{BeleznayKunimoto22}. Our in-situ halo set corresponds to a HJ occurrence rate of $0.17_{-0.10}^{+0.17}\%$ ($< 0.19\%$ assuming no detections), while our accreted set corresponds to an upper limit of $<0.56\%$ based on no detections.

The deficit of HJs relative to the Galactic disc is further underscored by a forward-model comparison, where we compare the number of planets we detected to what we would have expected assuming planets followed \textit{Kepler}-like planet abundances. Specifically, we simulated planets around stars in our halo sample following FGK star occurrence rates from \citet{Hsu19}. For each planet, we performed a Bernoulli draw where a planet has a probability of detection (a ``success'') based on its overall \textit{TESS}-based detection probability. We then repeated this forward-modelling process 100 times to get a sense of the spread in possible simulated detections. 

Assuming \textit{Kepler}-like planet abundances, we would have expected to find $10\pm3$ planets (mean $\pm$ standard deviation) in our halo sample, of which $7\pm3$ would be classified as HJs, yet we observed at most one. This stark discrepancy indicates a substantial suppression of close-in giant planets in the halo. By contrast, constraints on smaller planets ($2 < R_p < 8~R_{\oplus}$) are weak. Although \textit{Kepler}-like populations predict $3\pm2$ such planets, none are observed, but the low detection efficiency in this regime precludes strong conclusions.



\section{Discussion}\label{sec:discussion}

Our \textit{TESS} search yields no confirmed short-period planets and two HJ candidates, one of which appears highly grazing in the accreted subsample. 
Our occurrence constraints indicate that short-period HJ--sized objects are less common in the halo than in the thin disc, and for smaller planets we place upper limits extending into the sub-Saturn and sub-Neptune regimes. Taken together, the results sharpen the emerging picture that close-in planets, particularly large planets, are intrinsically rare in the old, metal-poor halo.

Our conclusions are consistent with prior null results around halo stars, while tightening constraints in several places. \citet{Boley21} used \textit{TESS} to set limits on HJs among halo dwarfs, finding a $1\sigma$ upper limit of $<0.18\%$, while \citet{Yoshida22} found $<0.52\%$ for metal-poor giants, both based on no detections. A recent \textit{TESS} survey of likely Sagittarius stream members likewise reported no planet detections and derived upper limits on HJ occurrence in that accreted halo structure \citep{Schap26}. Relative to previous studies, our work adds: (i) a larger, uniformly vetted dwarf sample matched to \textit{TESS}’s cadence across 93 sectors; (ii) the first HJ occurrence rate based on a planet candidate detection for halo stars ($0.13_{-0.07}^{+0.12}$\%); and (iii) the first occurrence-rate comparison between in-situ and accreted halo populations within a homogeneous, kinematically selected Milky Way halo sample based on actions-energies. This combination provides a more differentiated baseline against which to test formation channels at low metallicity.

\subsection{In-situ versus accreted halo}

Due to the lack of detections, we cannot conclude that there are significant differences in short-period occurrence between in-situ and accreted subsamples. 
A similar conclusion was also reached recently in estimating close binary fractions in halo subpopulations \citep{Bashi24}. If this parity holds to exoplanets as well, it suggests that the factors suppressing close-in planets in the halo operated similarly in the proto-Milky Way and in the dwarf progenitors subsequently accreted \citep[e.g.][]{Helmi18,Yuan20, Belokurov18, BelokurovKravtsov22, Thomas25}.

\subsection{Implications for planet formation}

In core accretion, dust surface density governs core growth, hence giant planets should be disfavoured at low metallicity because cores struggle to reach runaway before gas dispersal \citep{Pollack96, Mordasini12}. Our limits in an ancient, metal-poor population align with that expectation and with the broader metallicity dependence of close-in planets in the disc \citep[e.g.][]{FischerValenti05, Petigura18, KunimotoMatthews20, Zink23, Gan25}. Recent evidence for a metallicity cut-off even for super-Earth formation \citep{Boley24} reinforces the idea that both core assembly and envelope accretion become increasingly difficult in early, low-$Z$ regions. 

Two additional ingredients likely matter in the halo context. First, early protoplanetary disc conditions are important. Halo stars formed when discs were, on average, shorter-lived and exposed to harsher radiative environments; both effects work against giant-planet formation and inward migration \citep[e.g.,][]{HallattLee25}.
Second, stellar multiplicity may also play a role. Close-binary fractions in halo-like kinematics \citep{Bashi24} can inhibit disc growth, destabilise nascent planetary systems, and increase the rate of grazing, V-shaped photometric false positives. Disentangling multiplicity-driven selection from an intrinsic planet deficit is therefore essential.

For small planets, theoretical models allow a wider range of outcomes. Cores may still form at low $Z$ via locally enhanced dust/pebble fluxes, but retaining light envelopes is challenging when discs disperse quickly. Our present completeness below $\sim\!5\,R_\oplus$ cannot yet choose between a very low but non-zero occurrence and a true paucity. Coupling our framework to chemical-evolution predictions for the availability of solids and volatiles across Galactic components offers a promising route forward \citep{Cabral23,Nielsen23}.

\subsection{When in the Galaxy’s history should planets emerge?}

If metallicity and disc longevity are the primary levers, the first abundant close-in planets should appear near the onset of the high-$\alpha$ thick disc, as gas fractions, dust-to-gas ratios, and star-formation environments transitioned away from the most extreme early conditions \citep{BelokurovKravtsov22,BelokurovKravtsov23}. In that view, truly ancient, metal-poor halo stars would host few close-in planets, while high-$\alpha$ populations at intermediate metallicity would mark the rise in occurrence \citep[e.g.,][]{Bashi20}. A direct, age-resolved mapping, combining precise kinematics with robust stellar ages, is needed to test this timeline \citep[e.g.,][]{Sayeed25}.

\subsection{Follow-up and future prospects}

Our candidates, and especially the grazing event around an accreted host, warrant additional attention. While not included in our final planet candidate catalogue, the $\sim8.3~R_{\oplus}$ planet candidate orbiting the binary companion to TOI-5962 also warrants attention. Targeted follow-up, including ground-based photometry, high-resolution imaging, and a small number of RV measurements, will help confirm or reject their planetary nature. Future observations from ongoing \textit{TESS} sectors \citep{Ricker15} and, in the longer term, PLATO \citep{PLATO25}, and other dedicated RV surveys \citep{Aloisi25} could improve sensitivity and provide valuable constraints, offering a clearer picture of planet occurrence in the halo.


Our constraints are presently dominated by short periods and reduced completeness at small radii, with residual uncertainties from unresolved binarity and chemodynamical classification. 
Even at this stage, the estimated occurrences we obtain in both in-situ and accreted halo populations set strong boundary conditions on planet formation at the earliest, most metal-poor epochs of the Milky Way.

\section*{Acknowledgments}

This paper includes data collected with the \textit{TESS} mission, obtained from the MAST data archive at the Space Telescope Science Institute (STScI). Funding for the \textit{TESS} mission is provided by the NASA Explorer Program. STScI is operated by the Association of Universities for Research in Astronomy, Inc., under NASA contract NAS 5-26555.

This work has made use of data from the European Space Agency (ESA) mission \gaia~ (https:
//www.cosmos.esa.int/gaia), processed by the \gaia~ Data Processing and Analysis Consortium (DPAC, https://www.cosmos.esa.int/web/gaia/dpac/consortium). Funding for the DPAC has been provided by national institutions, in particular the institutions participating in the \gaia~ Multilateral Agreement.

MK acknowledges the support of the Natural Sciences and Engineering Research Council of Canada (NSERC), RGPIN-2024-06452. Cette recherche a été financée par le Conseil de recherches en sciences naturelles et en génie du Canada (CRSNG), RGPIN-2024-06452.

\section*{Data Availability}

The \textit{Gaia} DR3 data are publicly available at \url{https://gea.esac.esa.int/archive/}. QLP light curves and the CTL catalogue are available on MAST \url{https://archive.stsci.edu/hlsp/qlp}. 
The code and derived data products used in this paper will be made available upon reasonable request to the corresponding author.



\bibliographystyle{mnras}
\bibliography{Refs}




\bsp	
\label{lastpage}
\end{document}